\documentclass[%
 reprint,
 amsmath,amssymb,
 aps,
]{revtex4-1}

\usepackage{graphicx}
\usepackage{dcolumn}
\usepackage{bm}
\usepackage{amsfonts}

\begin{document}


\title{Kibble-Zurek mechanism in
a Dissipative Transverse Ising Chain}

\author{Hiroki Oshiyama}%
\email{hiroki.oshiyama.e6@tohoku.ac.jp}
\altaffiliation[Present address: ]{Graduate School of Information Sciences, Tohoku University, Sendai 980-8578, Japan}
\affiliation{%
 Department of Physics, Tohoku University, Sendai 980-8578, Japan
}%

\author{Sei Suzuki}
 \email{sei01@saitama-med.ac.jp}
 \affiliation{Department of Liberal Arts, Saitama Medical University,
 Moroyama, Saitama 350-0495, Japan}

\author{Naokazu Shibata}
 \email{shibata@cmpt.phys.tohoku.ac.jp}
\affiliation{%
 Department of Physics, Tohoku University, Sendai 980-8578, Japan
}%

\date{\today}

\begin{abstract}
We study the Kibble-Zurek mechanism in the transverse Ising chain coupled to
a dissipative boson bath, making use of a new numerical method with the infinite
time evolving block decimation combined with the discrete-time
path integral. We first show the ground-state phase diagram and confirm
that a quantum phase transition takes place in the presence of the 
system-bath coupling. Then we present the time dependence of the 
energy expectation value of the spin Hamiltonian and the scaling
of the kink density with respect to the time period over which the spin
Hamiltonian crosses a quantum phase transition. The energy of spins
starts to grow from the energy at the ground state of the full system
near a quantum phase transition. The kink density decays as
a power law with respect to the time period. These results confirm
that the Kibble-Zurek mechanism happens. We discuss the exponent
for the decay of the kink density in comparison with a theoretical
result with the quantum Monte-Carlo simulation. A comparison to an experimental
study is also briefly mentioned.
\end{abstract}

\maketitle


\section{\label{sec:Introduction}Introduction}

When a parameter of an isolated macroscopic system is varied
with a finite speed near a quantum phase transition, 
the system, departing from its ground state, develops into
a non-trivial out-of-equilibrium state with topological defects.
This spontaneous creation of inhomegeneity
during the time evolution near a quantum phase transition
is called the Kibble-Zurek mechanism (KZM)\cite{kibble_topology_1976,zurek_cosmological_1985}.
The most remarkable phenomenon of KZM is the universal scaling,
named as the Kibble-Zurek scaling (KZS), of the density of defects
with respect to the changing speed of the parameter.
Intensive studies in past decades have established that,
as far as a standard quantum phase transition of the second order
is concerned, the density of defects obeys a power law
with an exponent determined by critial exponents of the associated
quantum phase transition\cite{zurek_dynamics_2005,polkovnikov_universal_2005,polkovnikov_colloquium_2011}.

Compared to an isolated system, dissipative systems
have not been studied thoroughly so far. Previous studies on 
dissipative systems have been
limited to ideal free-fermion systems in one dimension with no mixing
between different momentum sectors \cite{patane_adiabatic_2008,patane_adiabatic_2009,nalbach_quantum_2015,dutta_anti-kibble-zurek_2016,keck_dissipation_2017,arceci_optimal_2018}.
However, it is no doubt that 
dissipation is crucially important to out-of-equilibrium states in
more real physical systems as well.
Therefore it is significant
to investigate KZM in dissipative quantum systems from more realistic
point of view.

Recently, a commercialized quantum annealing machine manufactured
by D-Wave Systems Inc. has received great attention. Quantum annealing
was proposed to solve combinatorial optimization problems
represented by the energy minimization of an interacting Ising-spin model\cite{kadowaki_quantum_1998,farhi_quantum_2001}.
Following the standard method of quantum annealing, one applies a strong
transverse field to interacting Ising spins at first and 
weaken it slowly with time. 
Then the spin state which is initialized as the trivial ground state
of the transverse field evolves adiabatically, and eventually reaches the target
ground state of the interacting Ising model when the transverse
field vanishes. The adiabatic time evolution 
is guaranteed when the square of the instantaneous energy gap above the ground state
is sufficiently large compared to the changing speed of the Hamiltonian\cite{kato_adiabatic_1950,messiah_quantum_1961}.
However, it has been known that a quantum phase transtition usually takes place
during quantum annealing and the instantaneous energy gap vanishes
at a transition\cite{sachdev_quantum_1999}. Therefore, quantum annealing is seen as a time
evolution across a quantum phase transition and its performance is
closely related to KZM.

D-Wave's machine
implements quantum annealing physically in its device. Although a lot of tests have
been done so far to characterize the performance of the machine, 
its mechanism has not been fully clarified. Several studies have shown
that the system embedded in D-Wave's machine is not
closed but open to a dissipative environment. D-Wave's machine is now believed to
implement quantum annealing in a dissipative system.
Thus KZM in a dissipative transverse Ising model is worth studying
 from the viewpoint of not only out-of-equilibrium statistical mechanics
but also revealing the mechanism of D-Wave's machine.

The difficulty in theoretically studying a dissipative quantum many-body system 
is that there have been few good method.
In order to study the time evolution of a dissipative quantum system,
the Born-Markov (BM) approximation is often used\cite{weiss_quantum_2012}. BM approximation
reduces the complexity of computation, but it involves an approximation
in an uncontrollable manner and one cannot improve accuracy systematically.
The Lindblad formalism, which is derived
with BM approximation, enables us to perform numerical computation
for large systems, i.e., tens of quantum spins for instance. 
The approximation in the Lindblad formalism replaces the
coupling to an environment with an abstract one and hence
some of the features of the original system are lost completely.
In the present work, we apply the infinite time evolving block
decimation (iTEBD) method to the dissipative transverse Ising chain
for the first time.
The dissipative transverse Ising chain is mapped to a double-layered
two-dimensional Ising model after tracing out the bosonic degrees of freedom. The difference from the dissipationless
model is that there are long-range interactions along the time axis.
The use of matrix product representation for the spin chain along
the time axis enables us to utilize the iTEBD method to the dissipative
model. This is essantially an extention of the method for finite system
that the present authors developed recently\cite{suzuki_quantum_2019}.
The present method does not rely on the BM approximation
and the approximation involved is controlled by parameters. With the present
method, one can study the time evolution as well as the thermal
equilibrium of the translationally invariant
infinite system without any finite-size effect.

Using our method, we study the equilibrium and out-of-equilibrium states
of the dissipative transverse Ising chain with 
the Caldeira-Leggett type of the system-bath coupling.
As shown in Sec. \ref{sec:Model}, this model cannot be mapped to
a free-fermion model and has not been studied so far in the context of
KZM. 
We first present the ground-state phase diagram, confirming that 
the quantum phase transition survives in a certain range of the
system-bath coupling strength. 
Then we investigate the scaling of the density of defects
associated with the change of the Hamiltonian linear in time across
a quantum phase transition. It is shown that the density of defects
scales as a power law of the changing speed of the transverse field, 
and that the exponent diminishes gradually with increasing the strength of the
system-bath coupling. The result is discussed in comparison to the
previous work. We finally comment on experiments of KZM in the
present model using D-Wave's machines.

This paper is organized as follows. We specify the dissipative
transverse Ising chain in the next section. In Sec. \ref{sec:QUAPI-iTEBD},
we describe our new method for equilibrium and out-of-equlibrium states
of the dissipative transverse Ising chain based on the discrete-time
path integral and iTEBD.
Section \ref{sec:Results} is devoted to the presentation of our results.
The ground-state phase diagram of the present model is given and 
KZS is discussed in detail there. This paper conludes with Sec. \ref{sec:Conclusion}.

\section{\label{sec:Model}Model}
We consider the dissipative transverse Ising chain described by the Hamiltonian
\begin{equation}
 H = H_{\rm S} + H_{\rm B} + H_{\rm int} ,
\end{equation}
where $H_{\rm S}$, $H_{\rm B}$ and $H_{\rm int}$ represent
the transverse Ising chain, the boson bath, and the interaction
between them, respectively, and given by
\begin{equation}
 H_{\rm S} = - J\sum_{j = 1}^{N-1}\sigma_{j}^{z}\sigma_{j+1}^z
  - \Gamma\sum_{j=1}^N\sigma_j^x ,
\end{equation}
\begin{equation}
 H_{\rm B} = \sum_{j=1}^N\sum_a \omega_a b_{j a}^{\dagger}b_{j a} ,
\end{equation}
\begin{equation}
 H_{\rm int} = \sum_{j=1}^N\sigma_j^z\sum_a\lambda_{a}\left(b_{j a}^{\dagger} + b_{j a}\right) ,
\label{eq:Hint}
\end{equation}
Here $\sigma_j^{c}$ ($c = x$, $z$) are the Pauli operators at site $j$,
$J$ and $\Gamma$ stand for the coupling strength of Ising spins and the transverse field,
respectively, 
and $b_{j a}^{\dagger}$ ($b_{j a}$) is the bosonic creation (annihilation)
operator at site $j$ with mode $a$.
We make $\hbar = 1$ throughout the paper.
We assume that the boson bath has the Ohmic spectral density
\begin{equation}
 \mathcal{J}(\omega) = \sum_a\lambda_a^2\delta(\omega - \omega_a)
  = \alpha\omega e^{-\omega/\omega_c} ,
\end{equation}
where $\alpha$ is the coupling strength between the spin system and the bath,
and $\omega_c$ is the cutoff energy of the bath spectrum which is chosen to be
larger than the other energy scales. 
The spin-boson interaction assumed in Eq.~(\ref{eq:Hint}) is the
many spin version of the Caldeira-Leggett model for the superconducting
flux qubit\cite{caldeira_quantum_1983,leggett_dynamics_1987}. The present model is the simplest and the most realistic
one to study the dissipative system in D-Wave's machine.

The presence of the system-bath coupling breaks the integrability of the
transverse Ising chain, $H_{\rm S}$. The nonintegrability of the present model
makes it difficult to study out-of-equilibrium states. 
We utilize the discrete-time path integral and infinite
time evolving block decimation (iTEBD) to study the infinitely 
large system.

\section{\label{sec:QUAPI-iTEBD}Method}

In this section, we describe the numerical method to compute the reduced density matrix and physical quantities. 

\subsection{Discrete-time path integral}

\subsubsection{\label{sec:statics}Thermal equilibrium}

We consider first the thermal equilibrium state of the full system.
The reduced density operator for the spin subsystem 
is given by
\begin{equation}
 \rho^{\rm eq}_{\rm S}(\beta) \equiv \frac{1}{Z}
\label{eq:Def_RDMeq}
 {\rm Tr}_{\rm B}\left(e^{-\beta H}\right) ,
\end{equation}
where $\beta$ is the inverse temperature, $Z$ is the partition function,
and ${\rm Tr}_B$ denotes the trace with respective to the bosonic degree of freedom.
The exponential operator in RHS is arranged, using the symmetric Trotter formula\cite{trotter_product_1959,suzuki_quantum_1986}, into
\begin{eqnarray}
 e^{-\beta H} &=& \left(e^{-\Delta\tau H_{\rm int}/2}e^{-\Delta\tau (H_{\rm S} + H_{\rm B})}
  e^{- \Delta\tau H_{\rm int}/2}\right)^M + O(\Delta\tau^2) \nonumber\\
 &=& e^{-\beta H_{\rm S}}e^{-\beta H_{\rm B}}
  e^{- \Delta\tau H_{\rm int}^{\rm I}(M\Delta\tau)/2}
  e^{- \Delta\tau H_{\rm int}^{\rm I}((M-1)\Delta\tau)} \nonumber\\
 && \cdots
  e^{- \Delta\tau H_{\rm int}^{\rm I}(\Delta\tau)}
  e^{- \Delta\tau H_{\rm int}^{\rm I}(0)/2} + O(\Delta\tau^2),
\label{eq:Trotter}
\end{eqnarray}
where $M$ denotes the Trotter number, $\Delta\tau = \beta/M$, and 
$H_{\rm int}^{\rm I}(l\Delta\tau)$ is the interaction picture of $H_{\rm int}$
defined by 
\begin{eqnarray}
 H_{\rm int}^{\rm I}(\tau) &\equiv& e^{\tau H_{\rm S}}e^{\tau H_{\rm B}}
H_{\rm int}e^{-\tau H_{\rm B}}e^{-\tau H_{\rm S}} \label{eq:IntPict}\\
 &=& \sum_{j=1}^N e^{\tau H_{\rm S}}\sigma_j^z e^{-\tau H_{\rm S}}
  \sum_a\lambda_a
  (b_{j a}^{\dagger}e^{\omega_a\tau} + b_{j a}e^{-\omega_a\tau}) . \nonumber
\end{eqnarray}
Using Eqs.~(\ref{eq:Trotter}) and (\ref{eq:IntPict}) in Eq.~(\ref{eq:Def_RDMeq}),
and inserting the completeness relations between exponential operators,
one obtains a path integral representation for the reduced density matrix.
After performing the Gaussian integrals with respect to bosonic degree of freedom,
the resultant path-integral formula of the reduced
density matrix is written as
\begin{eqnarray}
 &&\langle\boldsymbol{\sigma}_0 | \rho_{\rm S}^{\rm eq} (\beta = M\Delta\tau)
  |\boldsymbol{\sigma}_M\rangle \label{eq:QUAPI_imaginarytime}\\
 &&\approx C \sum_{\boldsymbol{\sigma}_1}\cdots\sum_{\boldsymbol{\sigma}_{M-1}}
 \prod_{j=1,N}\mathcal{F}^{\rm eq}(\sigma_{j,0},\cdots,\sigma_{j,M}) 
 \nonumber\\
 && ~~~ \times \prod_{l=0}^M \prod_{j=1}^{N-1}
   \left\{\mathcal{B}^{\rm eq}(\sigma_{j,l},\sigma_{j+1,l})\right\}^{1 - \frac{1}{2}\delta_{l,0}
   - \frac{1}{2}\delta_{l,M}}  , \nonumber
\end{eqnarray}
where $|\boldsymbol{\sigma}_l\rangle$ denotes the eigenbasis of $\sigma_j^z$ 
satifying
$\sigma_j^z|\boldsymbol{\sigma}_l\rangle = \sigma_{j,l}|\boldsymbol{\sigma}_l\rangle$,
and $C$ is a normalization factor. $\mathcal{B}^{\rm eq}$ and $\mathcal{F}^{\rm eq}$ are
defined by
\begin{equation}
 \mathcal{B}^{\rm eq}(\sigma_{j,l},\sigma_{j+1,l}) 
  = \exp\left[\Delta\tau
	 \sigma_{j,l}\sigma_{j+1,l}
	\right] ,
\end{equation}
and
\begin{eqnarray}
 &&\mathcal{F}^{\rm eq}(\sigma_{j,0},\cdots,\sigma_{j,M}) \\
 &&= \exp\Biggl[
	  \Upsilon\sum_{l=0}^{M-1}\sigma_{j,l}\sigma_{j,l+1} \Biggr. \nonumber\\
 &&~~~\Biggl.
	  + \alpha\Delta \tau^2\sum_{M\geq l > m\geq 0}
	  \mathcal{K}((l-m)\Delta\tau)\sigma_{j,l}\sigma_{j,m} 
	 \Biggr] ,\nonumber
\end{eqnarray}
respectivly, where
\begin{equation}
 \Upsilon = \frac{1}{2}\log \coth \Gamma\Delta\tau ,
\end{equation}
and
\begin{equation}
 \mathcal{K}(\tau) = \int\frac{\cosh(\beta\omega/2 - \tau\omega)}{\sinh\beta\omega/2}
  \mathcal{J}(\omega)d\omega .
\end{equation}
Note that Eq.~(\ref{eq:QUAPI_imaginarytime}) represents the partition function of a 
$(1+1)$ dimensional Ising model
with the long-range interaction
in imaginary time. 

\subsubsection{Out of equilibrium}

We next consider the time-dependent dissipative transverse Ising chain
with
$H_{\rm S}(t) = -J(t)\sum_{i=1}^{N-1}\sigma_j^z\sigma_{j+1}^z - \Gamma(t)\sum_{j=1}^N\sigma_j^x$,
and assume the initial state being a product state between a pure spin state
denoted by $|\Psi_0\rangle$ and the thermal equilibrium of the boson bath:
\begin{equation}
 \rho(0) = |\Psi_0\rangle\langle\Psi_0| \bigotimes 
  \frac{e^{-\beta H_{\rm B}}}{Z_{\rm B}} ,
\end{equation}
where $\beta$ is the inverse temperature of the boson bath and
$Z_{\rm B} = {\rm Tr}_{\rm B}e^{-\beta H_{\rm B}}$.

Let $U(t)$ be the time-evolution operator that satisfies
$i\frac{d}{dt}U(t)=H(t)U(t)$. The reduced density operator is written as
\begin{equation}
 \rho_{\rm S}(t) = {\rm Tr}_{\rm B}\left(U(t)\rho(0)U^{\dagger}(t)\right) .
\label{eq:RDM_realtime}
\end{equation}a
In spite of the time dependence in the Hamiltonian, an algebra with
the symmetric Trotter decomposition and Gaussian integrals similar to those
used in the previous subsection yield the path-integral formula \cite{makarov_path_1994,suzuki_quantum_2019}:
\begin{eqnarray}
 &&\langle\boldsymbol{\sigma}_M|\rho_{\rm S}(M\Delta t)|\boldsymbol{\sigma}'_M\rangle 
    \label{eq:QUAPI_realtime}\\
&&\approx D\sum_{\boldsymbol{\sigma}_0,\boldsymbol{\sigma}'_0}\cdots
 \sum_{\boldsymbol{\sigma}_{M-1},\boldsymbol{\sigma}'_{M-1}}
 \prod_{j=1}^N\mathcal{F}(\sigma_{j,0},\sigma'_{j,0},\cdots,\sigma_{j,M},\sigma'_{j,M}) 
 \nonumber\\
 &&~~~ \times \left[\prod_{l=0}^M\prod_{j=1}^{N-1}
	       \mathcal{B}_l(\sigma_{j,l},\sigma'_{j,l},\sigma_{j+1,l},\sigma'_{j+1,l}) \right]
 \langle\boldsymbol{\sigma}_0|\Psi_0\rangle\langle
 \Psi_0|\boldsymbol{\sigma}'_0\rangle \nonumber
\end{eqnarray}
with
\begin{eqnarray}
 &&\mathcal{B}_l(\sigma_{j,l},\sigma'_{j,l},\sigma_{j+1,l},\sigma'_{j+1,l})
  \\
 &&
  = \exp\left[
	 i\Delta t J_l\left(\sigma_{j,l}\sigma_{j+1,l} - \sigma'_{j,l}\sigma'_{j+1,l}\right)
	\right] , \nonumber
\end{eqnarray}
and
\begin{widetext}
\begin{eqnarray}
 &&\mathcal{F}(\sigma_{j,0},\sigma'_{j,0},\cdots,\sigma_{j,M},\sigma'_{j,M}) 
  \label{eq:out-of-equilibrium_F}\\
 &&= \exp\left[
       \sum_{j=1}^N \sum_{l=1}^M
       \left(\gamma_l\sigma_{j,l}\sigma_{j,l-1} + 
	\gamma_l^{\ast}\sigma'_{j,l}\sigma'_{j,l-1} \right)
      \right. \nonumber\\
 &&~~~
  \left. + \alpha\Delta t^2 \sum_{j=1}^N
  \left\{K(0)\sum_{l=0}^M\sigma_{j,l}\sigma'_{j,l} 
 - \sum_{M\geq l > m\geq 0}
  \left(\begin{array}{cc}
   \sigma_{j,l} & \sigma'_{j,l} \\
  \end{array}\right)
  \left(\begin{array}{cc}
   K((l-m)\Delta t) & -K^{\ast}((l-m)\Delta t)\\
	 -K((l-m)\Delta t) & K^{\ast}((l-m)\Delta t)
	\end{array}\right)
  \left(\begin{array}{c}
   \sigma_{j,m}\\
	 \sigma'_{j,m}
	  \end{array}\right)\right\}
\right] , \nonumber
\end{eqnarray}
\end{widetext}
where $D$ is a normalization factor. The coupling parameters between Ising spins are given by
\begin{equation}
 J_l = \left\{\begin{array}{ll}
	\frac{J(\Delta t) + 3 J(0)}{8}  & \mbox{for $l=0$} \vspace{1ex}
\\
	       \frac{3J(M\Delta t) +  J((M-1)\Delta t)}{8}  & \mbox{for $l=M$} \vspace{1ex}\\
	       \frac{J((l+1)\Delta t) + 6 J(l\Delta t) + J((l-1)\Delta t)}{8} & \mbox{otherwise}
	      \end{array}
\right. ,
\end{equation}
\begin{equation}
 \gamma_l = \frac{1}{2}\log
  \left(-i\cot\frac{\Gamma(l\Delta t) + \Gamma((l-1)\Delta t)}{2}\Delta t\right)
\end{equation}
and
\begin{equation}
 K(t) = \int\frac{\cosh(\beta\omega/2 - i\omega t)}{\sinh \beta\omega/2}
  \mathcal{J}(\omega)d\omega .
\end{equation}
We remark 
that two types Ising variables, $\sigma_{j,l}$ and $\sigma'_{j,l}$,
are included in Eq.~(\ref{eq:QUAPI_realtime}). 
This is because the reduced density operator in Eq.~(\ref{eq:RDM_realtime}) involves
the time-forward operator $U(t)$ and the time-backward one $U^{\dagger}(t)$.
The former results in $\sigma_{j,l}$ and the latter in $\sigma'_{j,l}$ in Eq.~(\ref{eq:QUAPI_realtime}).
Hence the effective Hamiltonian $\mathcal{H}$ represents coupled double layers of
two-dimensional Ising models with nearest neighbor interactions in space
and long-range interactions in time. Since the strength of the long-range interaction 
$K(t)$ decays with $t$, we set a cutoff $t_c$ in the numerical simulation
so that $K(t) = 0$ for $t > t_c$.

\subsection{iTEBD}

In this subsection, we briefly describe how to compute the
reduced density matrix, Eq.~(\ref{eq:QUAPI_realtime}),
for an out-of-equilibrium state.
The method mentioned here can be easily applied to
the thermal equilibrium state.
We hereafter use a shortened notation $S_{j,l}$ for $(\sigma_{j,l}, \sigma'_{j,l})$
in such a way that $S_{j,l} = \frac{1}{2}(1 - \sigma_{j,l}) + (1 - \sigma'_{j,l})$.

Using the singular value decomposition (SVD),
one can write Eq.~(\ref{eq:out-of-equilibrium_F}) in the form of a matrix product as
\begin{eqnarray}
 &&\mathcal{F}(S_{j,0},\cdots,S_{j,M}) \label{eq:MPS_F}\\
 &&=  \sum_{\eta_{1},\cdots,\eta_{M}}
  \phi^{(0)S_{j,0}}_{\eta_{1}}\phi^{(1)S_{j,1}}_{\eta_{1}\eta_{2}}
  \cdots\phi^{(M-1)S_{j,M-1}}_{\eta_{M-1}\eta_{M}}
  \phi^{(M)S_{j,M}}_{\eta_{M}} . \nonumber
\end{eqnarray}
The factor $\mathcal{B}_l$ can be represented, in a similar manner, 
as
\begin{eqnarray}
 &&\prod_{j=1}^{N-1}\mathcal{B}_l(S_{j,l},S_{j+1,l}) \label{eq:MPS_B}\\
 && = \sum_{\zeta_1,\cdots,\zeta_N}\Xi^{(1,l)S_{1,l}}_{\zeta_1}
  \Xi^{(2,l)S_{2,l}}_{\zeta_1,\zeta_2}\cdots
  \Xi^{(N,l)S_{N,l}}_{\zeta_{N-1},\zeta_N} .
  \nonumber
\end{eqnarray}
Finally, we write 
$\langle\boldsymbol{\sigma}_0|\Psi_0\rangle\langle\Psi_0|\boldsymbol{\sigma}'_0\rangle$
as
\begin{eqnarray}
 &&\langle\boldsymbol{\sigma}_0|\Psi_0\rangle\langle\Psi_0|\boldsymbol{\sigma}'_0\rangle
  \label{eq:MPS_Psi}\\
 && = \sum_{\mu_1,\cdots,\mu_N}
  \Lambda^{(1,0)S_{1,0}}_{\mu_1}\Lambda^{(2,0)S_{2,0}}_{\mu_1,\mu_2}\cdots
  \Lambda^{(N,0)S_{N,0}}_{\mu_{N-1},\mu_{N}} .\nonumber
\end{eqnarray}
Strictly speaking, the largest matrix dimensions of 
$\phi^{(l)S}$ and $\Lambda^{(j,l)S}$
increase exponentially with $M$ and $N$, respectively.
However, as is the case with the ground-state wave function in 
a one-dimensional quantum system, one can choose the basis
of matrices so that the necessary information is preserved
with keeping the largest matrix dimension constant in $M$ or $N$ \cite{white_density_2005,white_real-time_2004}.

Let us now consider the infinite system ($N\to\infty$) with the translational invariance. In order to perform the trace over the spin degrees of freedom in Eq. (\ref{eq:QUAPI_realtime}), we exploit the iTEBD algorithm \cite{vidal_classical_2007,orus_infinite_2008}. 
This algorithm performs the trace with respect to $\{S_{j,l}\}$ for all $j$ from $l = 0$
to $l = M-1$ iteratively.
After the $M$ times iteration, we obtain the infinite matrix product state (iMPS) representation of $\rho_{\rm S}(t)$, from which we can calculate physical quantities. 

Hereafter, we write the $l$-th level of iMPS representation generated from the $l-1$ level as $\psi^{(l)\eta_{j,l}}_{\mu_l,\mu'_l}$, where $\eta_{j,l}$ denotes the matrix index for 
$\phi^{(l)S_{j,l}}$ in Eq.~(\ref{eq:MPS_F}), and $\mu_l$ and $\mu'_l$ denote matrix indices. We denote the largest matrix dimension of $\phi^{(l)S_{j,l}}$ and $\psi^{(l)\eta_{j,l}}$ by $\chi_t$ and $\chi_s$, respectively.

For $l=1$, the iMPS is given as
\begin{eqnarray}
\psi^{(1)\eta_{j,1}}_{\mu_1,\mu'_1}=\sum_{S_{j,0}}\phi^{(0)S_{j,0}}_{\eta_{j,1}}\Xi^{(j,0)S_{j,0}}_{\zeta_{j,0},\zeta_{j+1,0}}
\Lambda^{(0)S_{j,0}}_{\mu,\mu'}.
\end{eqnarray}
where we define the new matrix indices as $\mu_1=(\zeta_1,~\mu)$ and $\mu'_1=(\zeta_1,~\mu')$. 
For $1 \leq l \leq (M-2)$,
the $(l+1)$-th level of iMPS is calculated by using the $l$-th one as
\begin{eqnarray}
\psi^{(l+1)\eta_{j,l+1}}_{\mu_{l+1},\mu'_{l+1}}=\sum_{S_{j,l},\eta_{j,l}}\phi^{(l)S_{j,l}}_{\eta_{j,l},\eta_{j,l+1}}
\Xi^{(j,l)S_{j,l}}_{\zeta_{j,l},\zeta_{j+1,l}}
\psi^{(l)\eta_{j,l}}_{\mu_{l},\mu'_{l}}
.
\end{eqnarray}
where $\mu_{l+1}=(\zeta_{j,l},~\mu_{l})$ and $\mu'_{l+1}=(\zeta_{j,l},~\mu_{l}')$.
For each step, one can keep the matrix dimension up to $\chi_s$ by means of the truncation algorithm in iTEBD.
The fianl iMPS is calculated as
\begin{eqnarray}
\psi^{S_{j,M}}_{\mu_{M+1},\mu'_{M+1}}=\sum_{\eta_{j,M}}\phi^{(M)S_{j,M}}_{\eta_{j,M}}
\Xi^{(j,M)S_{j,M}}_{\zeta_{j,M},\zeta_{j+1,M}}
\psi^{(M)\eta_{j,M}}_{\mu_{M},\mu'_{M}}
.
\end{eqnarray} 

Thus the reduced density matrix is given in terms of iMPS as
\begin{equation}
\langle\boldsymbol{\sigma}_M|\rho_{\rm S}(M\Delta t)|\boldsymbol{\sigma}'_M
\rangle
= {\rm tr}[\cdots\psi^{S_{j,M}}\psi^{S_{j+1,M}}\cdots]
,
\end{equation}
where ${\rm tr}$ denotes the matrix trace.
The expectation values of physical quantities are obtained in terms of $\psi^{S}$
as follows.
Redefining $\psi^{S}$ and $\tilde{\psi} \equiv \psi^{S=0} + \psi^{S=3}$ in such a way that the largest eigenvalue of $\tilde{\psi}$ is unity, the magnetization is given by
\begin{eqnarray}
 {\rm Tr}_{\rm S}\left[\sigma^{c}\rho_{\rm S}(M\Delta t)\right]
  &=& \lim_{N\to\infty}{\rm tr}\left[
  \Bigl(\sum_{\sigma,\sigma'}\langle\sigma|\sigma^c|\sigma'\rangle
  \psi^{S(\sigma,\sigma')}\Bigr)(\tilde{\psi})^N\right] \nonumber\\
 &=&\vec{v}^{\,t}_{\rm L}  \left(\sum_{\sigma,\sigma'}\langle\sigma|\sigma^c|\sigma'\rangle
  \psi^{S(\sigma,\sigma')}\right)
 \vec{v}_{\rm R} ,
\label{eq:iTEBD_single_spin}
\end{eqnarray}
where $\vec{v}^{\,t}_{\rm L}$ and $\vec{v}_{\rm R}$ are the left and right
eigenvector of $\tilde{\psi}$ for the unit eigenvalue that satisfy the normalization
condition $\vec{v}^{\,t}_{\rm L}\vec{v}_{\rm R} = 1$.
The correlation function is given by
\begin{eqnarray}
 &&{\rm Tr}_{\rm S}
  \left[
   \sigma_0^c\sigma_r^c\rho_{\rm S}(M\Delta t)\right] 
  \label{eq:iTEBD_two_spins}\\
 &&= \vec{v}^{\,t}_{\rm L}
  \left(
   \sum_{\sigma,\sigma'}
   \sum_{\tau,\tau'}
   \langle\sigma|\sigma_0^c|\sigma'\rangle
   \langle\tau|\sigma_r^c|\tau'\rangle
   \psi^{S(\sigma,\sigma')}
   (\tilde{\psi})^{r-1}
   \psi^{S(\tau,\tau')}
   \right)
   \vec{v}_{\rm R} . \nonumber
\end{eqnarray}
Equations (\ref{eq:iTEBD_single_spin}) and
(\ref{eq:iTEBD_two_spins}) are used to compute expectations of physical quantities.

\section{\label{sec:Results}Results}

In order to study properties of the dissipative transverse Ising chain in
and out of equilibrium, we introduce a parameter $s$ for the system Hamiltonian,
so that $J = s$ and $\Gamma = 1 - s$, as
\begin{equation}
 H_{\rm S}(s) = - s\sum_{i} \sigma^z_i\sigma^z_{i+1} - (1 - s) \sum_i \sigma^x_i .
\end{equation}
This Hamiltonian reduces to the decoupled spins in a transverse field when $s = 0$,
while it corresponds to the Ising model when $s = 1$. 
In the absence of coupling to the bath, 
the ground state of $H_{\rm S}(s)$ is an ordered one for $\frac{1}{2} < s < 1$,
a disoredered one for $0\leq s < \frac{1}{2}$, and quantum critical at $s = \frac{1}{2}$.
The Hamiltonian of the full system is given by 
$H = H_{\rm S}(s) + H_{\rm B} + H_{\rm int}$.
We will focus on the zero-temperature phase diagram of the full system in the next subsection
and then move to the time evolution in Sec. \ref{sec:Dynamics}.

\subsection{\label{sec:PhaseDiag}Phase diagram}

\begin{figure}[t]
 \begin{center}
  \includegraphics[width=7cm, bb = 0 0 369 275]{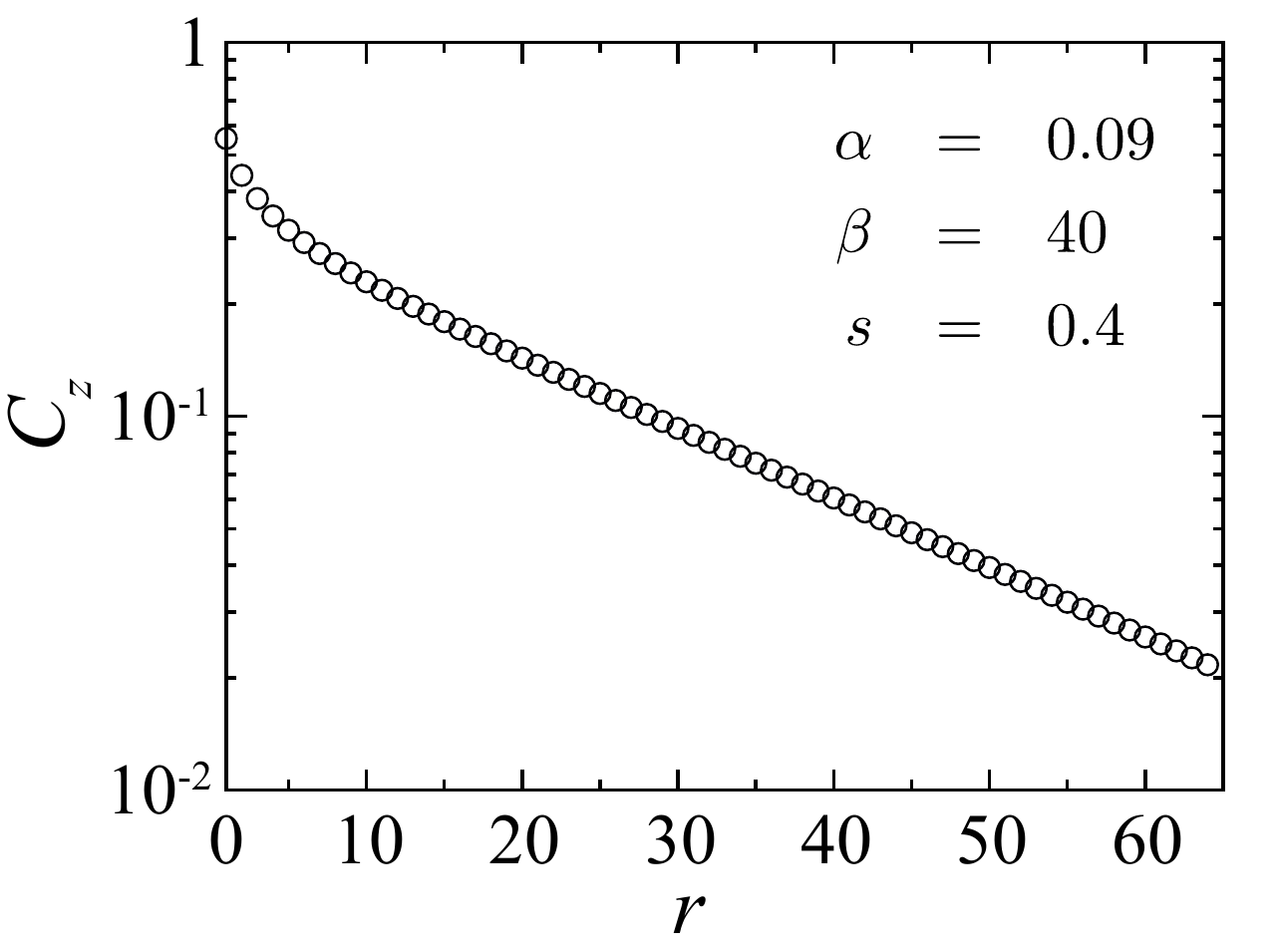}
\\
  \includegraphics[width=7cm, bb = 0 0 365 289]{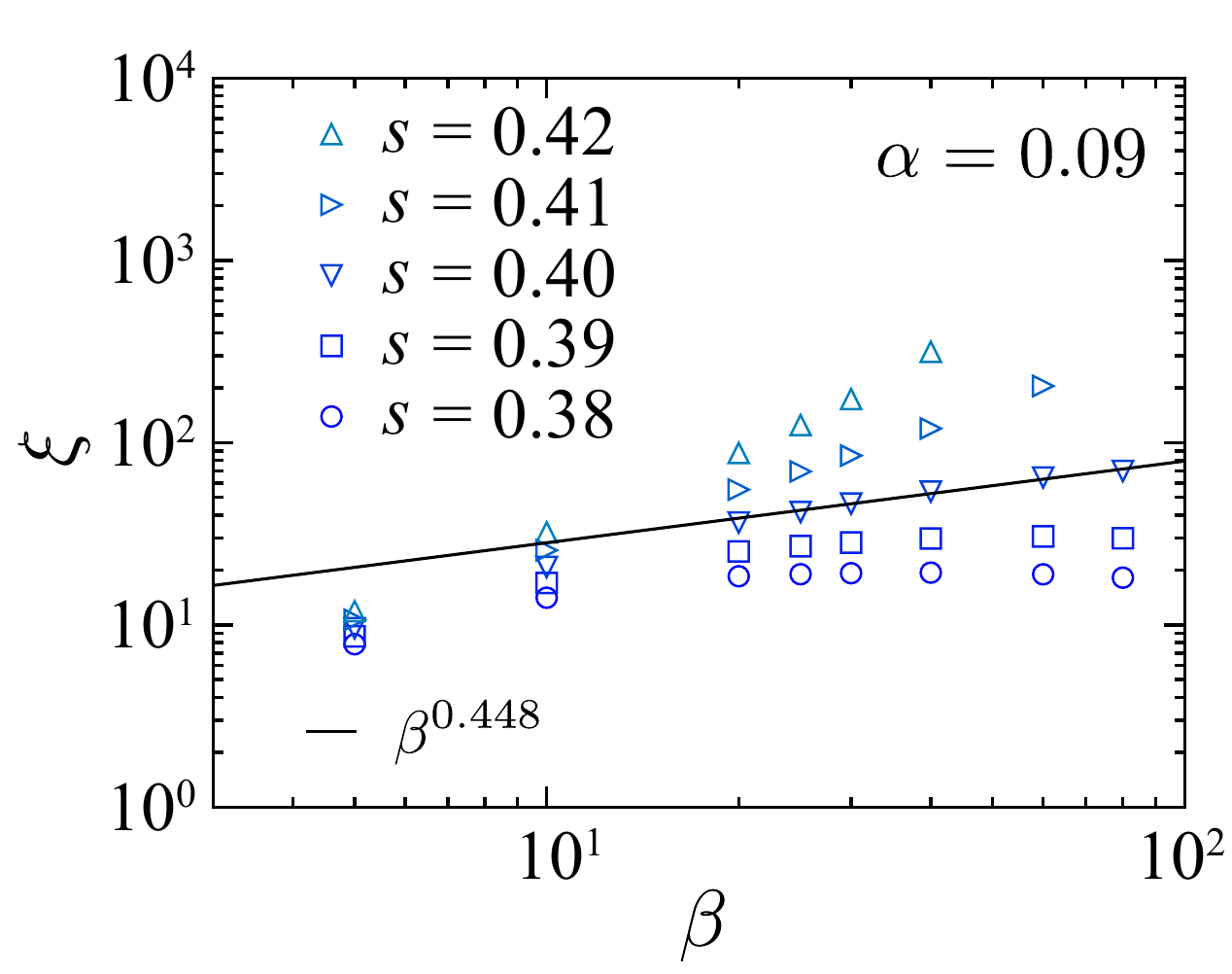}
 \end{center}
\caption{(a) The longitudinal-spin correlation function of the dissipative
transverse Ising chain with $s = 0.4$ and $\alpha = 0.09$ at $\beta = 40$.
The correlation function decays exponentially with separation between spins.
One can extract the correlation length from this behavior.
(b) The correlation length $\xi$ as a function of the inverse temperatue $\beta$
for $\alpha = 0.09$ and several $s$. One can find the location $s_c$ of the quantum 
critical point from the power-law growth of $\xi$ for
sufficiently large $\beta$. In fact, one finds $s_c\approx 0.40$ for $\alpha = 0.09$
from the figure.
The parameters for numerical simulations were chosen as
$\Delta\tau = 0.05$, $\chi_t = 128$, $\chi_s = 128$, and $\omega_c$ = 5.}
\label{fig:Correlation}
\end{figure}

Consider that the full system is in the thermal equilibrium with the inverse temperature
$\beta$. The reduced density operator is given by Eq.~(\ref{eq:Def_RDMeq}) and
the correlation function with respect to the longitudinal spin is given by
\begin{equation}
 C^z(r) = {\rm Tr}_{\rm S}\left[\sigma_0^z\sigma_r^z\rho_{\rm S}^{\rm eq}(\beta)
       \right] .
\end{equation}
It is expected that, since the longitudinal spin is disordered in the equilibrium
at a finite temperature, the correlation function 
decays exponentially for sufficiently large $r$ as
\begin{equation}
 C^z(r) \sim e^{-r/\xi} ,
\label{eq:Cr_ansatz}
\end{equation}
where $\xi$ represents the correlation length.
Figure \ref{fig:Correlation}(a) demonstrates an exponential decay of 
$C^z(r)$ obtained by the iTEBD simulation.

Based on the ansatz of Eq.~(\ref{eq:Cr_ansatz}), 
we estimate the correlation length from the correlation function
by
\begin{equation}
 \xi = - \lim_{r\to\infty}\frac{r}{\log C^z(r)} .
\end{equation}
With lowering the temperature, the correlation length obeys the scaling
relation to $\beta$ depending upon the property of the ground state.
When the ground state is disordered, the correlation length becomes constant
with respect to the temperature. When the ground state is ordered, on the other hand,
the correlation length diverges exponentially as
$\xi\sim e^{a\beta}$ as $\beta\to\infty$ with a constant $a$.
Finally, a quantum critical ground state bears a power-law scaling,
$\xi\sim \beta^{1/z}$, where $z$ is the dynamical critical exponent.
We show results on $\xi$ for several $s$ in Fig.~\ref{fig:Correlation}(b). 
One can see that $\xi$ grows faster than a power law with $\beta$ for sufficiently 
large $s$'s, while it looks converging for smaller $s$'s. At a certain critical
value of $s$ in between, $\xi$ behaves as a power law of $\beta$, from which
one can determine the location of the quantum critical point $s_c$ for a fixed
system-bath coupling $\alpha$.

\begin{figure}[t]
\begin{center}
 \includegraphics[width=7cm, bb = 0 0 348 276]{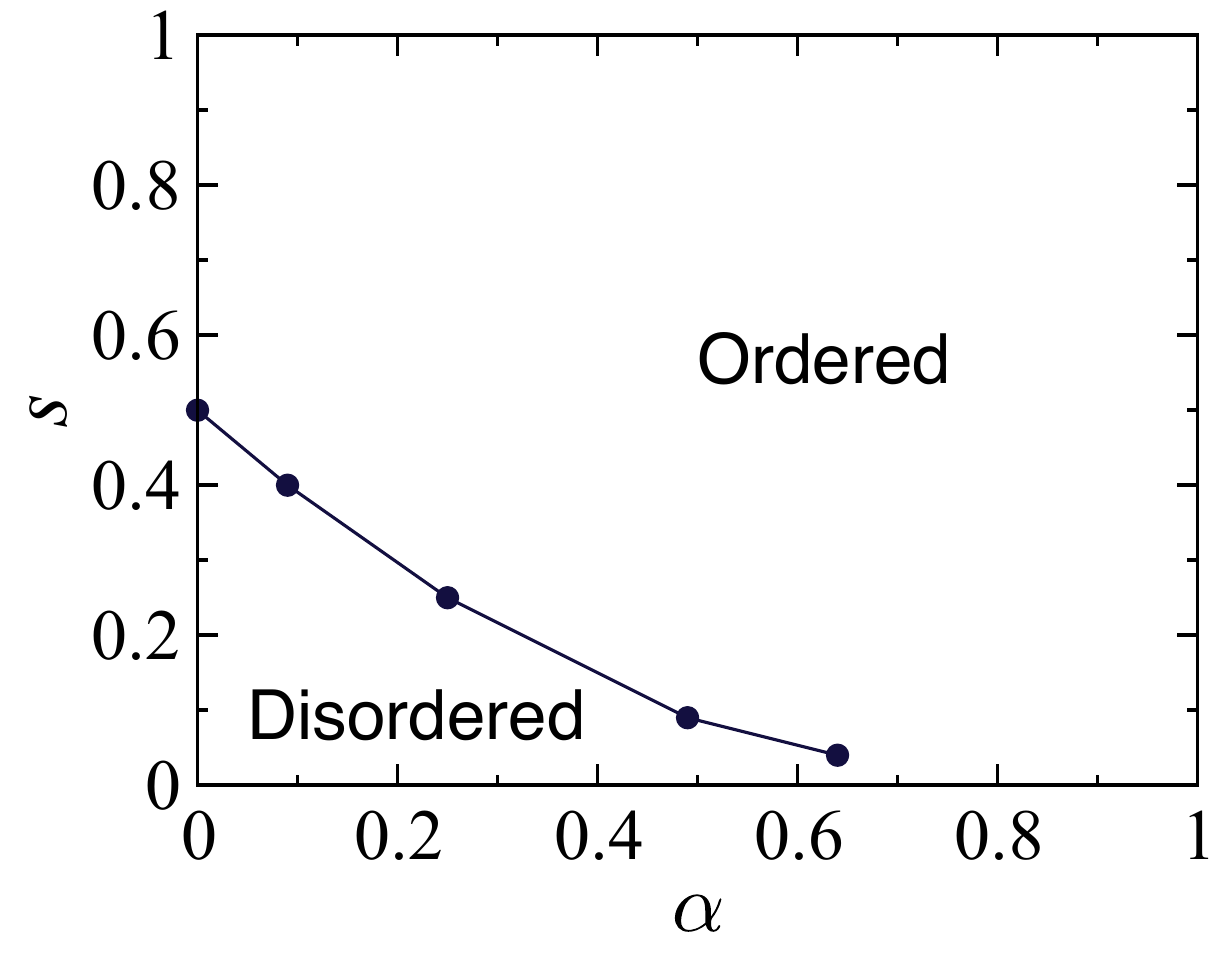}
\end{center}
\caption{Ground-state phase diagram of the dissipative transverse Ising chain.
With increasing the system-bath coupling, the ordered phase extends down to
a smaller critical value of $s$.}
\label{fig:GS_PhaseDiagram}
\end{figure}

Thus we show the ground-state phase diagram of the dissipative
Transverse Ising chain in Fig. \ref{fig:GS_PhaseDiagram}. 
Note that, at $\alpha = 0$, the transverse Ising chain is isolated from
the bath and the critical point $s = \frac{1}{2}$ is exactly known.
One finds that the critical point declines with increasing the system-bath
coupling and the ordered phase extends down to a smaller $s$.
Although our numerical method becomes unreliable for $\alpha > 0.64$, previous
studies have revealed that when $s = 0$ an order-disorder transition
takes place at $\alpha \approx 0.625$ for $\omega_c = 10$ 
and at larger $\alpha$ for smaller $\omega_c$
\cite{leggett_dynamics_1987,bulla_numerical_2003,werner_phase_2005,Strathearn_efficient_2018}. 
This implies that the critical line should
terminate at $\alpha$ larger than 0.625 and $s = 0$. Our results for $\omega_c = 5$
are perfectly consistent to the previous studies.

Quantum critical properties of the dissipative transverse Ising chain
have been studied earlier by Werner et al. using the quantum Monte-Carlo
(QMC) simulation for an effective model \cite{werner_phase_2005}. 
They have found that 
the critical exponents $(\nu,z)$ associated with the quantum phase transition
change from $(1,1)$ to $(0.63, 2.0)$ by introducing an infinitesimal
system-bath coupling $\alpha$ and do not vary on the critical line.
As for our simulation, the exponent $z$
can be evaluated from the power law of $\xi$ when $s$ is fixed at a quantum 
critical point. The result is in fact $\frac{1}{z}\approx 0.46$ for $\alpha = 0.09$, which
is in reasonable agreement with $\frac{1}{2}$ from the QMC simulation.

It should be noted that the parameters involved in the QMC
simulation are defined for an effective model with the unit Trotter
imaginary-time step, i.e., $\Delta \tau = 1$.
Therefore there is not necessarily a quantitative consistency
among the phase diagrams given in ref.\cite{werner_phase_2005}
and Fig.~\ref{fig:GS_PhaseDiagram}.

\subsection{\label{sec:Dynamics}Time evolution across a quantum phase transition}

\begin{figure}[t]
 \begin{center}
  \includegraphics[width=8cm, bb = 0 0 357 221]{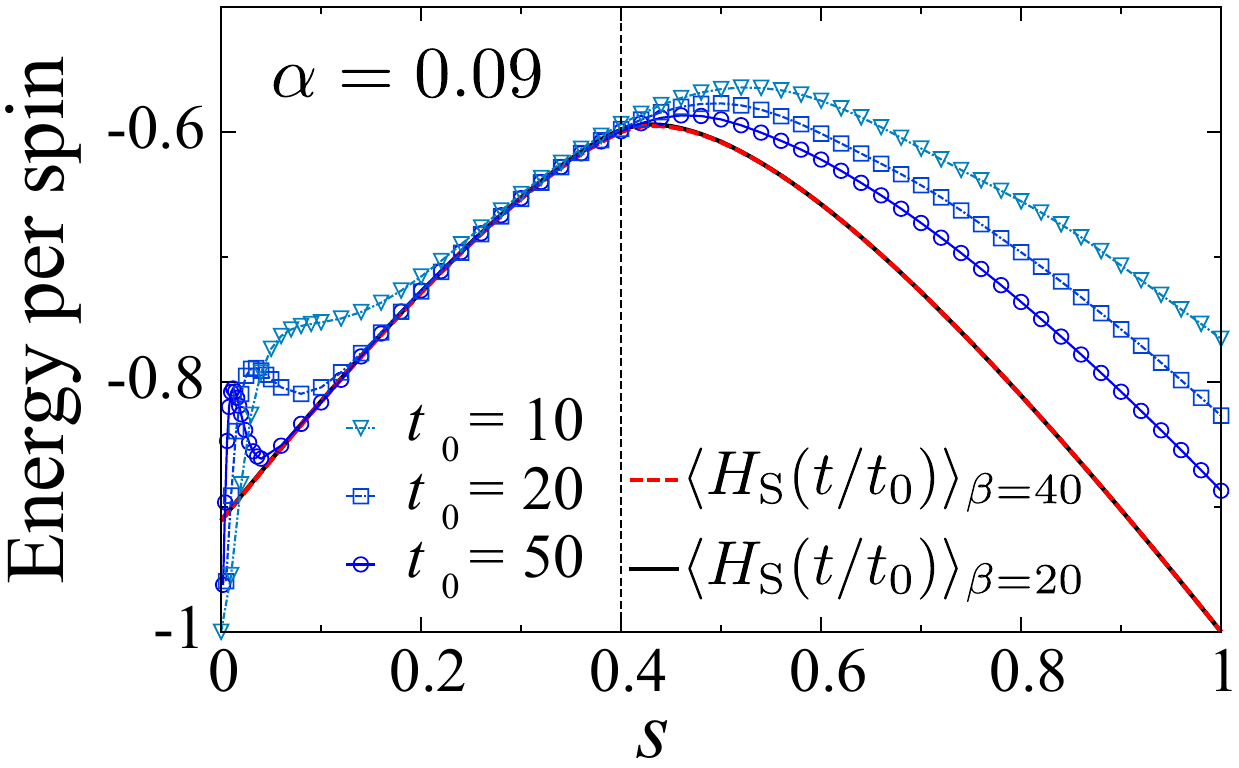}
 \end{center}
\caption{Time dependence of the energy expectation value of the spin
Hamiltonian $H_{\rm S}(t/t_0)$ for various $t_{0}$ and the system-bath
coupling fixed at $\alpha = 0.09$. The solid lines show the instantaneous
energy expectation values of the thermal equilibria, $\rho_{\rm S}^{\rm eq}(\beta)$, 
with $\beta = 40$ and $20$. Since the difference between these two lines is
invisible, we consider that the solid line represents the ground state.
As for the time evolution, the initial state is assumed to be
the product state of the ground state of the spin system and that
of the boson system. 
The dashed vertical line shows the location of the quantum phase transition.
After an initial relaxation, the state follows the ground state but
starts to deviate from the ground state near the quantum phase transition.
The parameters for numerical simulations were chosen as
$\Delta t = 0.05$, $\chi_t = 128$, $\chi_s = 128$, and $\omega_c = 5$.}
\label{fig:s_Energy}
\end{figure}
Having obtained the ground state phase diagram, we next focus on the time evolution across
a quantum phase transition.
Let us consider the time-dependent Hamiltonian, 
$H(t) = H_{\rm S}(t/t_0) + H_{\rm B} + H_{\rm int}$ with
\begin{equation}
 H_{\rm S} (t/\tau) = - \frac{t}{t_0}\sum_{i}\sigma_i^z \sigma_{i+1}^z
  - \left(1 - \frac{t}{t_0}\right)\sum_i\sigma_i^x ,
\end{equation}
where time $t$ moves from $0$ to $t_0$. The parameter $t_0$
controls the changing speed of $H_{\rm S}$. The larger $t_0$ is,
the slower $H_{\rm S}$ varies.
As seen in the previous subsection, the present model involves a quantum phase transition.
When the system-bath coupling $\alpha$ is not so large, The system undergoes a quantum
phase transition at a certain normalized time $s = t/t_0$ for sufficiently large $t_0$. See Fig.~\ref{fig:GS_PhaseDiagram}.

Suppose that the full system is initially in the ground state. 
On approaching a quantum phase transition,
we expect that the full system is excited because the characteristic time
of the instantaneous ground state exceeds the time scale $t_0$ of
Hamiltonian's variation.
This is illustrated by Fig.~\ref{fig:s_Energy}, where the energy expectation 
value of the spin Hamiltonian $H_{\rm S}(t/t_0)$ is plotted as a 
function of the normalized time $t/t_0$ for various $t_0$.
We have assumed in our simulation that the full system is initialized as the product state of 
the ground states of the spin system and the boson system.
Although this is not the true ground state of the full system, we find
that the state relaxes into the ground state of the full system in an initial
period. After that, the full system keeps the ground state up to a time near the
quantum phase transition, where the state starts to deviate from the ground state.

It has been known that the excitation during time evolution near a quantum phase transition
is well explained by KZM. 
According to the phenomenological theory of KZM, the state after passing
a quantum phase transition involves topologial defects with a characteristic
length $\hat{\xi}$ which is scaled by $t_0$ as $\hat{\xi}\sim t_0^{\nu/(z\nu + 1)}$,
where $\nu$ and $z$ are the critical exponents \cite{zurek_dynamics_2005,polkovnikov_universal_2005}. This universal scaling
relation is KZS.
In our model, the topological defects is identified by the kinks
between neighboring spins in the final state. The kink density is then defined by
\begin{equation}
 n = \frac{1}{2}\left[1 - {\rm Tr}\left(\sigma_0^z\sigma_1^z\rho_{\rm S}(t_0)\right)\right] .
\end{equation}
Using KZS, this should scale as
\begin{equation}
 n \sim \hat{\xi}^{-1} \sim t_0^{ - \nu/(z\nu + 1)} .
\end{equation}

\begin{figure}[t]
 \begin{center}
  \includegraphics[width = 8cm, bb = 0 0 376 273]{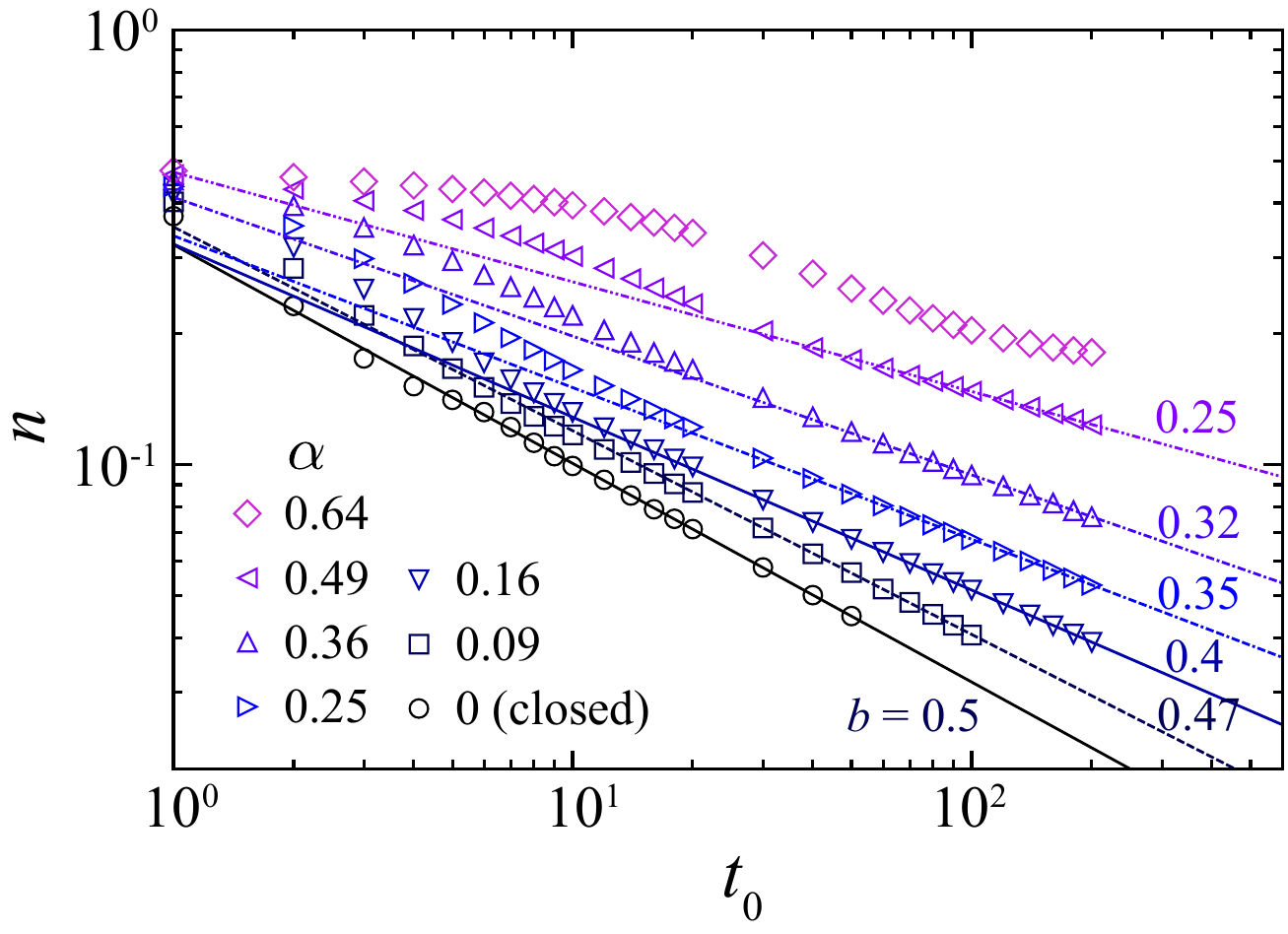}
 \end{center}
\caption{Kink density as a function of $t_0$ for various $\alpha$. 
Solid lines show fitting by a power law $t_0^{-b}$
of the rightmost eight data for each $\alpha$. The exponents $b$ thus obtained
are shown on the figure. With increasing $\alpha$ from 0, $b$ decreases
from $0.5$.
The parameters for numerical simulations were chosen as
$\Delta t = 0.05$, $\chi_t = 128$, $\chi_s = 128$, and $\omega_c$ = 5.
}
\label{fig:Eexc_scaling}
\end{figure}

Figure \ref{fig:Eexc_scaling} shows results of our simulation on 
the kink density $n$ as a function
of $t_0$. The kink density decays monotonically with $t_0$ and
obeys the power law $n\sim t_0^{-b}$ approximately for large $t_0$.
The exponent $b$ drawn by fitting the data
decreases from 0.5 to around 0.25
with strengthening the system-bath coupling $\alpha$ from 0 up to
0.49, beyond which it is uncertain from our data
whether a power law is valid or not.

As mentioned above, the exponents $\nu$ and $z$ are known to
be $\nu = z = 1$ for $\alpha = 0$ by the exact solution of the transverse Ising chain,
whereas they are estimated as $\nu \approx 0.63$ and $z \approx 2.0$ for $\alpha > 0$ 
by the QMC simulation \cite{werner_phase_2005}.
These values predict that $b = 0.5$ for $\alpha = 0$ and $b \approx 0.28$
for $\alpha > 0$.
Comparing them to our result, the fact that $b$ is reduced by coupling
spins to the bath is shared by the theory and our simulation.
However, although the theory predicts a discontinuous change of $b$ from 0.5
to 0.28 by an infinitesimal $\alpha$, our results implies a continuos
change of $b$. Moreover, the smallest value of $b$ by our simulation,
which is obtained for $\alpha = 0.49$, is slightly smaller than 0.28.
These discrepancies might be attributed to the range of $t_0$.
Although our method of simulation is not able to access 
$t_0 > 200$, the slope of $n$ might be closer to that of $t_0^{-0.28}$
if one has much larger $t_0$.

Recently, KZM of the transverse Ising chain was studied experimentally 
using D-Wave's machines. As mentioned in Sec. \ref{sec:Introduction},
the physical system in D-Wave's machine can be regarded as the dissipative
transverse Ising model and hence D-Wave's machine should serve as a quantum
simulator of the present model. In fact, it has been reported in ref.\cite{bando_probing_2020}
that the kink density decays as $t_0^{-b}$ with 
$b \approx 0.20$ by the device in NASA and $\approx 0.34$ by the low-noise one
in Burnaby. We note that the range of $t_0$ investigated in these experiments
is roughly 4-digit larger than ours. Anyway, the fact that the exponent
decreases from 0.5 is common to
the experiment, our simulation, and the theory with the QMC simulation. 
As for the numerical value, the comparison between the experiment
and our simulation is difficult, since the strength of the system-bath coupling
is unknown for D-Wave's machines. Nontheless, $b = 0.34$ obtained experimentally
lies in the range from 0.5 for $\alpha = 0$ to $0.25$ for $\alpha = 0.49$
by our simulation. Therefore our results are consistent with the
experimental result in ref.\cite{bando_probing_2020}.

\section{\label{sec:Conclusion}Conclusion}

We investigated KZM of a dissipative transverse Ising chain.
First we presented the ground-state phase diagram and confirmed that 
a quantum phase transition survives even at a finite coupling strength
between the spin system and the boson bath. We next showed 
the energy expectation value of the spin subsystem as a function of time
when the spin Hamiltonian varies linearly in time 
from that of the decoupled spins in the transverse field to that of the Ising chain
and when the full system is initialized as the product state of the ground states
for spins and bosons.
We found that the energy of the spin subsystem relaxes into the
value at the ground state of the full system soon and starts to deviate from it
at around the quantum critical point.
This result implies that KZM takes place in our model.
We finally showed the kink density as a function of the time period
with which the spin Hamiltonian varies. We observed that 
the kink density decays as a power law with an exponent $b$ and that
$b$ decreases with increasing the system-bath coupling.
This fact partly agrees with a theoretical deduction with a QMC result that
$b$ changes from 0.5 to 0.28 by an introductin of the system-bath coupling, though
the values of $b$ do not always agree. 
Our results on the kink density, on the other hand, agrees well with an
 experimental result by D-Wave's machine.
Geting better understanding on $b$ remains to study. To this end, quantum simulators
like D-Wave's machine would be useful.

We used a new numerical algorithm in the present work, which is a combination
of the discrete-time path integral representation for the reduced density matrix and iTEBD. 
This method was for the first time applied to a dissipative quantum
many-body model. So far there had been no method to study out-of-equilibrium
states of such a model with large system size and without the Born-Markov approximation.
The present study will pave the new way to the study of
out-of-equilibrium physics in a dissipative quantum many-body system.

The authors thank H. Nishimori and Y. Susa for valuable discussions
and Y. Bando, M. Ohzeki, F. J. G\'omez-Ruiz, A. del Campo, and D. A. Lidar
for collaboration on a related experimental project.

\bibliography{KZM_DTIM}


\end{document}